\documentclass[a4paper,11pt]{article}
\pdfoutput=1 % if your are submitting a pdflatex (i.e. if you have
\usepackage{jcappub} % for details on the use of the package, please
\usepackage{graphicx}
\usepackage{caption}
\usepackage{subcaption}                     
\usepackage[T1]{fontenc} % if needed

\title{Imprints of non-standard Dark Energy and Dark Matter Models on the 21cm Intensity Map Power Spectrum}

\author[a,b,c]{Isabella P. Carucci,\footnote{Corresponding author.}}
\author[c]{Pier-Stefano Corasaniti,}
\author[a,b,d]{Matteo Viel}

\affiliation[a]{SISSA-International School for Advanced Studies, Via Bonomea 265, 34136 Trieste, Italy}
\affiliation[b]{INFN sez. Trieste, Via Valerio 2, 34127, Trieste, Italy}
\affiliation[c]{LUTH, Observatoire de Meudon, 5 Place Janssen, 92195 Meudon, France}
\affiliation[d]{INAF, Osservatorio Astronomico di Trieste, Via Tiepolo 11, 34143 Trieste, Italy}

\emailAdd{ipcarucci@sissa.it}

\abstract{We study the imprint of non-standard dark energy (DE) and dark matter (DM) models on the 21cm intensity map power spectra from high-redshift neutral hydrogen (HI) gas. To this purpose we use halo catalogs from N-body simulations of dynamical DE models and DM scenarios which are statistically indistinguishable from the standard Cold Dark Matter model with Cosmological Constant ($\Lambda$CDM) using currently available cosmological observations. We limit our analysis to halo catalogs at redshift $z=1$ and $2.3$ which are common to all simulations. For each catalog we model the HI distribution by using a simple prescription to associate the HI gas mass to N-body halos. We find that the DE models leave a distinct signature on the HI spectra across a wide range of scales, which correlates with differences in the halo mass function and the onset of the non-linear regime of clustering. In the case of the non-standard DM model significant differences of the HI spectra with respect to the $\Lambda$CDM model only arise from the suppressed abundance of low mass halos. These cosmological model dependent features also appear in the 21cm spectra. In particular, we find that future SKA measurements can distinguish the imprints of DE and DM models at high statistical significance.}
  
\begin{document}
\maketitle
\flushbottom

\section{Introduction} \label{sec:intro} 
Cosmological observations provide strong evidence that the Universe is dominated by dark invisible components. On the large scales cosmic expansion is currently undergoing an accelerated phase triggered by an exotic unknown dark energy (DE) component \cite{Riess1998,Perlmutter1999}, while on smaller scales an invisible dark matter (DM) drives the formation and evolution of the visible structures in the Universe (see e.g. \cite{Tegmark2004,Clowe2006,Massey2007}). In the standard cosmological model, a cosmological constant ($\Lambda$) in Einstein's equation of General Relativity can account for the DE phenomenon. Together with the cold dark matter (CDM) paradigm, which assumes DM to be composed of collisionless particles, the standard $\Lambda$CDM model has been tremendously successful at reproducing the cosmological observations thus far available \cite{Planck2015}. However, the puzzling conceptual issues raised by the existence of a small non-vanishing $\Lambda$ (see e.g. \cite{Carroll2001}) and the lack of detections of dark matter particle candidates in Earth laboratories pose fundamental questions on the origin of the invisible components. 

A variety of scenarios have been proposed in a vast literature to account for dark energy and dark matter, whether in the form of non-Standard Model fields \cite{Wetterich1988,Ratra1988,Caldwell1998} or a manifestation of deviations from Einstein's gravity (see e.g. \cite{Clifton2012}). However, to date none of the proposed models provide a self-consistent theoretical explanation of these phenomena. On the other hand, there is a widespread consensus that the study of the clustering of matter in the Universe may reveal key insights on their origin. In fact, measurements of the growth rate of structures may test whether DE is dynamical or a strictly static phenomenon as expected in the case of $\Lambda$. This is because by affecting the late time cosmic expansion, dark energy alters the rate of gravitational collapse of dark matter density fluctuations. Similarly, the CDM paradigm can be tested through measurements of the abundance and density profiles of DM structures at small scales since in non-standard DM models dissipative effects suppress the amplitude of small density fluctuations with respect to the standard cosmological model predictions. 

In the upcoming years a new generation of experiments in cosmology will provide accurate measurements of the clustering of matter in the Universe across an unprecedented range of scales and redshifts. Surveys such as Large Synoptic Survey Telescope (LSST) \cite{LSST} and Euclid \cite{Euclid} will detect billions of galaxies, while the Square Kilometer Array\footnote{http://skatelescope.org} (SKA) will map the distribution of neutral hydrogen from unresolved galaxies up to the high-redshift Universe.  

Observations of the 21cm line emission from cosmic neutral hydrogen are of particular interest since these will probe the distribution of matter in the redshift range $1\lesssim z \lesssim 3$, which extends to the matter dominated era where the effect of DE on the cosmic expansion is subdominant. By testing such a transitional regime, measurements of the 21cm power spectrum at large scales are expected to provide constraints on DE that are complementary to those obtained by other cosmic probes (see e.g. \cite{Wyithe2007,Chang2008,Bull2015,Kohri2017}), and constraints on the nature of DM particles  \cite{Sitwell2014,Sekiguchi2014,Carucci2015} that are also complementary to those obtained by investigating small scales observables \cite{Weinberg2015}.

Several studies have modelled the 21cm emission mostly using prediction from linear perturbation theory. However, HI is hosted in DM halos which are biased tracers of the DM density field, thus the use of numerical simulations may provide a more accurate description of the HI clustering. Moreover, numerical N-body simulation studies have also shown that DE alters the non-linear clustering of matter at small scales in a way that depends on the DE dynamics or lack thereof (see e.g. \cite{Casarini2009,Alimi2010,Jenning2010,Casarini2010,baldi2010,baldi2011}). For these reasons, it is timely to perform a forecast analysis of the 21cm signal of non-standard DE and DM scenarios using N-body simulations and test the sensitivity of SKA measurements to the different cosmological model predictions.

Here, we use large volume N-body simulations from the Dark Energy Universe Simulations (DEUS) database\footnote{http://www.deus-consortium.org/deus-data/} to derive prediction of the HI clustering in different DE models. In the case of the non-standard DM scenarios we use high-resolution simulations of axions and late-forming dark matter models presented in \cite{Corasaniti2017}, thus extending the analysis on the Warm Dark Matter scenario presented in \cite{Carucci2015}. We show that due to the fundamentally different way in which DE and DM models impact the non-linear regime of gravitational collapse, their signatures on the 21cm power spectrum remain distinct and can be differentiated by future SKA measurements. The paper is organised as the following: in Sec.~\ref{HImodel} we illustrate how we distribute neutral hydrogen in the simulations that we later present in Sec.~\ref{models}, together with the non-standard DE and DM models for which they are run. In Sec.~\ref{HIPk} we present the HI power spectra of the different cosmologies, we compare them and we discuss some numerical effects that enter into play. The 21cm signal results are shown in Sec.~\ref{SKA}, together with an estimation of the error with which the SKA telescope will be able to measure them. Finally, we summarise and draw our conclusions in Sec.~\ref{conclusions}.

\section{Neutral Hydrogen Distribution Model}\label{HImodel}
\subsection{N-body Halo Based Approach}
We model the neutral hydrogen distribution as in \cite{Bagla2010,Carucci2015,Carucci2017}. We assume that HI is confined in DM halos with a mass proportional to that of the host halo mass. Using this simple prescription we assign the neutral hydrogen mass to halos from N-body simulation catalogs. The spatial distribution of the resulting HI cloud catalog is then converted into 21cm maps. For each halo of mass $M$ at redshift $z$ the mass of neutral hydrogen is modelled as:
\begin{equation}\label{mhimodel}
M_{\rm HI}(M,z)=\begin{cases}
f M^{\alpha}\,\,\,\,\,\,{\rm if}\,\,M\ge M_{\rm min}\\
0\,\,\,\,\,\,\,\,\,\,\,\,\,\,\,\,{\rm otherwise}
\end{cases}
\end{equation}
where $\alpha=3/4$ consistently with results from hydrodynamic simulation studies \cite{Villaescusa2015,Villaescusa2016}, $f$ is a calibration parameter that depends on the cosmic abundance of neutral hydrogen and $M_{\rm min}$ is the minimum halo mass containing HI. The latter accounts for the fact that there is a neutral hydrogen density threshold below which the gas becomes fully ionised and unable to effectively self-shield from UV-radiation. As pointed out in \cite{Bagla2010}, stability arguments suggest that gas in halos with circular velocities  $v_{\rm circ}\gtrsim 60$ km s$^{-1}$ may undergo gravitational collapse to form stars, which corresponds to halos with a virial mass larger than
\begin{equation}
M\gtrsim 10^{10}\,M_{\odot}\left(\frac{v_{\rm circ}}{60\,{\rm km s^{-1}}}\right)^3\left(\frac{1+z}{4}\right)^{-1.5}.
\end{equation}
On the other hand, simulation analyses have shown that neutral hydrogen can reside even in smaller mass halos \cite{Pontzen2008}. As an example, using smoothed particle hydrodynamic simulations implemented with various baryon feedback models the authors of \cite{Villaescusa2016} have found that HI is hosted in halos with $v_{\rm circ}\gtrsim 25$ km s$^{-1}$. However, this was inferred by extrapolating the results on scale below the mass resolution of the simulations. Hence, a conservative cut-off is usually considered at $v_{\rm circ}\approx 25-30$ km s$^{-1}$, corresponding to a minimum halo mass $M_{\rm min}\sim 10^9$ M$_{\odot}$ at $z=3$. Given the large uncertainties on the exact value of $M_{\rm min}$, a larger mass cut can also be assumed, eventually never exceeding the mass of halos containing the most massive galaxies ($\sim 10^{11}$ M$_{\odot}$), since in the low redshift Universe groups and clusters of galaxies do not host large quantities of neutral hydrogen. As we will see our choice of $M_{\rm min}$ is mainly dictated by the mass resolution of the N-body simulation catalogs available to us.

The calibration parameter $f$ is set such that total amount of HI in the halo catalog reproduces the observed cosmic abundance. The cosmic HI density is given by
\begin{equation}
\Omega_{\rm HI}(z)=\frac{1}{\rho_c}\int_0^{\infty}\frac{dn}{dM}(M,z)\,M_{\rm HI}(M,z)dM,
\end{equation}
where $\rho_c$ is the present critical density and $dn/dM$ the halo mass function. Using Eq.~(\ref{mhimodel}) we obtain the following relation:
\begin{equation}
f=\frac{\Omega_{\rm HI}L^3\rho_c}{\sum_{i=0}^{N_{\rm halo }}M_i^{\alpha}\Theta(M_i-M_{\rm min})},
\end{equation}
where $L$ is the simulation box-lenght, $\Theta(x)$ is the Heaviside step function and $N_{\rm halo}$ is the number of halos in the catalog. Notice that the value of $\Omega_{\rm HI}$, only sets the overall amplitude of the 21cm signal and not the scale dependence of the HI distribution. Hence, though poorly constrained by observations, the exact value of $\Omega_{\rm HI}$ will not affect our analysis as we are interested in relative differences of the 21cm power spectrum predicted by different models. We set the value to $\Omega_{\rm HI}=10^{-3}$ consistently with observational results at $3\lesssim z\lesssim 5$ presented in \cite{Noterdaeme2009,Pochaska2009}.

It is important to stress that as we assign the HI mass to the center-of-mass of the halos without modelling its distribution within halos we expect the HI power spectrum to be valid only on scales approximately larger then the virial radius of the most massive halos in the simulation catalogs.

\subsection{21cm Emission Model}
Having assigned the HI mass to each halo with comoving (center-of-mass) position $\textbf{x}$, the redshift-space location of the HI cloud is given by
\begin{equation}
{\bf s}={\bf x}+\frac{1+z}{H(z)}{\bf v}_{\rm los}({\bf x}),
\end{equation}   
where $z$ is the redshift of the halo catalog considered, $H(z)$ is the Hubble function and ${\bf v}_{\rm los}$ is the component of the halo perculiar velocity along the line-of-sight. From the cloud distribution in redshift space one can compute through a standard cloud-in-cell algorithm the HI density $\rho_{\rm HI}({\bf s})$ and finally derive the brightness temperature fluctuation \cite{Mao2012}:
\begin{equation}
\delta T_b({\bf s})=\overline{\delta T_b}(z)\left[\frac{\rho_{\rm HI}(\textbf{s})}{\bar{\rho}_{\rm HI}}\right],
\end{equation}
where $\bar{\rho}_{\rm HI}$ is the HI mean density and 
\begin{equation}
\overline{\delta T_b}(z)=1571.05 \,\Omega_{\rm HI}h^2\sqrt{\frac{0.015(1+z)}{\Omega_m h^2}}\,{\rm mK}.
\end{equation}
The power spectrum of the 21cm intensity maps is then obtained by computing $P_{21cm}(k)=\langle\delta T_b({\bf k})\delta T_b^{*}({\bf k'})\rangle$. 

\section{Cosmological Models \& N-body Simulations}\label{models}
We use a set of simulations from the DEUS database of flat dark energy models. These consist of a standard cosmological model with cosmological constant ($\Lambda$CDM-W5) and two quintessence scenarios with dynamical equation of state as given by the scalar field evolution in a Ratra-Peebles \cite{Ratra1988} (RPCDM-W5) and supergravity \cite{SUGRA} (SUCDM-W5) self-interacting potentials respectively. As discussed in \cite{Alimi2010} the cosmological parameters of these models have been calibrated such as to reproduce within $1\sigma$ the cosmic microwave background power spectra from WMAP-5 observations \cite{WMAP5} and the luminosity distances from SN-Ia measurements \cite{SN} (see Table~\ref{table1}). 

\begin{table}[t]
\centering
\begin{tabular}{|c|c|c|c|c|}
\hline\hline
{\rm Model} & $\Omega_m$ & $\sigma_8$ &  $w_0$ & $w_a$ \\
\hline
$\Lambda$CDM-W5& 0.26 & 0.80 & -1 & 0 \\
\hline
RPCDM-W5& 0.23 & 0.66 & -0.87 & 0.08 \\
\hline
SUCDM-W5& 0.25 & 0.73 & -0.94 & 0.19 \\
\hline\hline
\end{tabular}
\caption{\label{table1} Cosmological model parameters of realistic DE models calibrated against WMAP-5 and SN Ia observations. The other cosmological parameters are set to $\Omega_b=0.04$, $h=0.72$, $n_s=0.96$. Notice that $w_0$ and $w_a$ the DE equation of state parameters of the Linder-Chevalier-Polarski parametrization \cite{Linder,Chevalier} best-fitting the time evolution of the quintessence-field equation of state.}
\end{table}

These "realistic" models are characterised by small differences in the large linear scale clustering which are amplified at small scales by the onset of the non-linear regime of gravitational collapse \cite{Alimi2010}. In particular, the quintessence models exhibit DM density power spectra in the range $0.1\lesssim k\, [{\rm Mpc}\,h^{-1}]\lesssim 1$ and $z\lesssim 2$ that are lower than the $\Lambda$CDM prediction with deviations as large as $20-40\%$. This is because the DE dynamics alters the cosmic expansion during the late matter dominated era by causing a less decelerated expansion than in the standard $\Lambda$CDM case. Consequently matter density fluctuations grow less efficiently than in $\Lambda$CDM which leads to a lower level of clustering \cite{Alimi2010} and halo abundances \cite{Courtin2011}. As we are interested in modelling the HI cloud distributions using N-body halo catalogs, we use data from the DEUS simulations with the largest available mass resolution. These have a $162$ Mpc $h^{-1}$ box-length and contain $1024^{3}$ particles (corresponding to mass particle resolution of $m_p=2.5\cdot10^{8}$ M$_{\odot}$ $h^{-1}$). To limit numerical systematic errors we only retain halos with more than 100 particles corresponding to a minimum halo mass of the catalogs of $M_{\rm min}^{162}=2.5\cdot 10^{10}$ M$_{\odot}$ $h^{-1}$. We identify the halos in each simulation box using the FoF halo finder \cite{pFoF}.

\begin{figure}[t]
\centering 
\includegraphics[width=.85\textwidth]{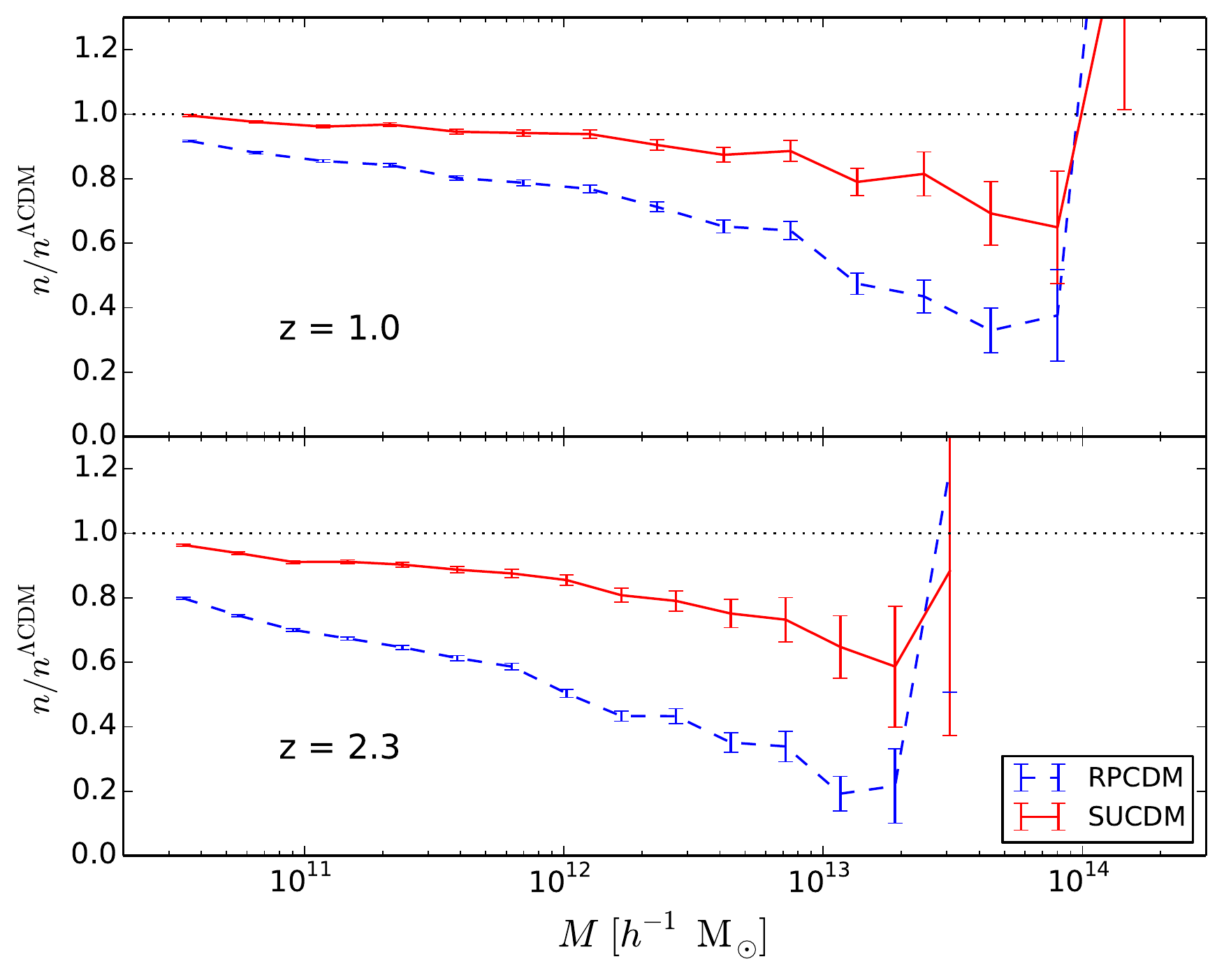}
\caption{\label{fig1} Ratios of the halo mass function of each quintessence model (RPCDM-W5 in blue solid line and SUCDM-W5 in red dotted) over the reference $\Lambda$CDM-W5 model at $z=1$ (top panel) and $2.3$ (bottom panel). Error bars represent Poisson errors.}
\end{figure}

In Fig.~\ref{fig1} we plot the ratio of the halo mass function of the quintessence model simulations relative to the reference $\Lambda$CDM-W5 at $z=1$ (top panel) and $2.3$ (bottom panel) respectively. As expected the dynamical DE models exhibit lower halo abundances with respect to the $\Lambda$CDM with increasing deviations as function of mass and redshift. Notice also that the RPCDM-W5 model has larger deviations ($\sim 20-80\%$) than SUCDM-W5 ($\sim 5-40\%$) which is consistent with expectations from the cosmic expansion history and linear growth rate of these models. 

\begin{figure}[t]
	\centering 
	\includegraphics[width=.85\textwidth]{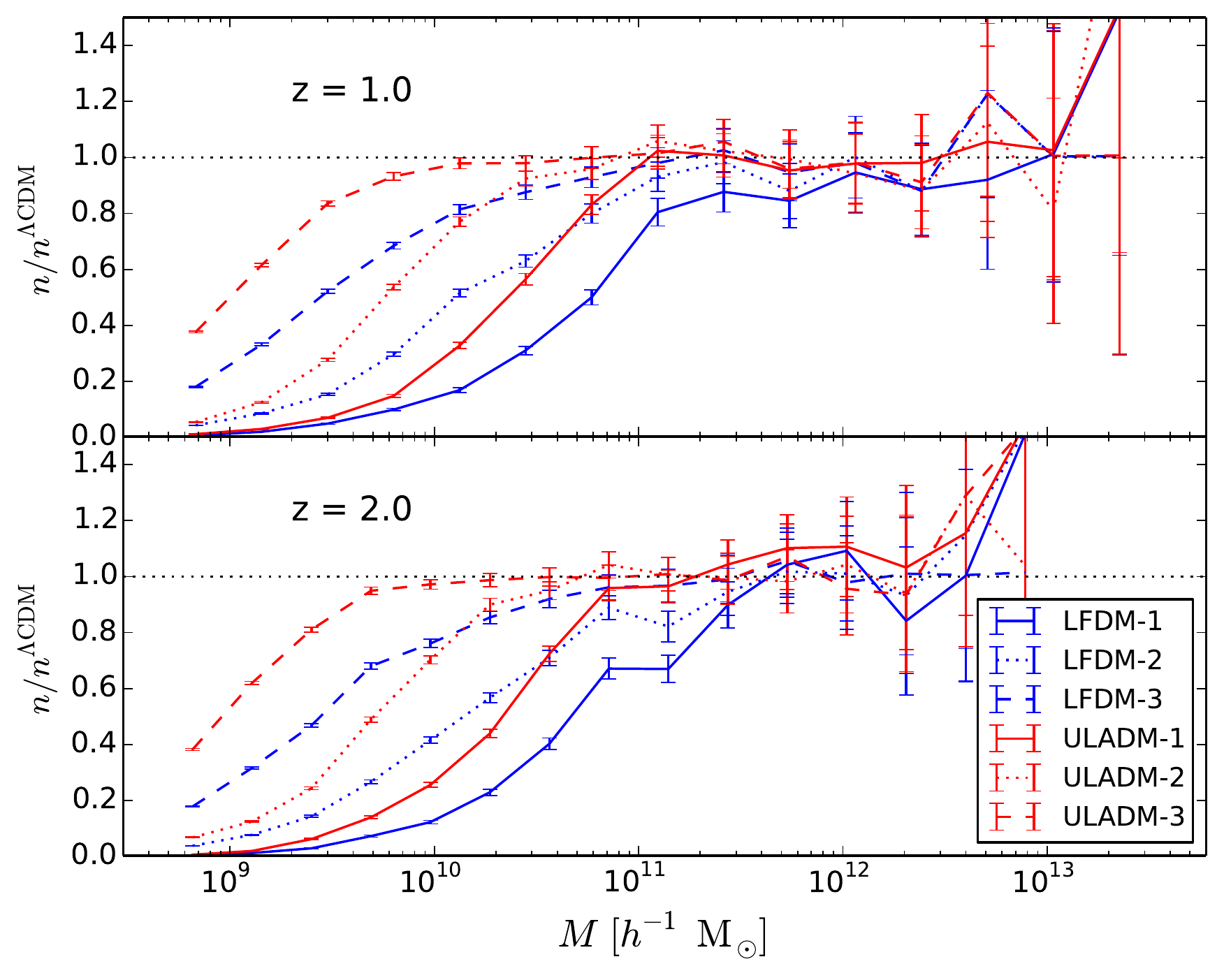}
	\caption{\label{fig2} Ratios of the halo mass function of the non-standard DM models relative to the reference $\Lambda$CDM scenario at $z=1$ (top panel) and $2$ (bottom panel) respectively. Error bars are represent Poisson errors.}
\end{figure}

In the case of the non-standard DM models, we consider N-body simulations of $27.5$ Mpc $h^{-1}$ box-length with $1024^3$ particles of non-standard DM scenarios presented in \cite{Corasaniti2017}. These consist of ultra-light axion DM models (\cite{Marsh2016} for a review) with axion mass $m_a=1.56\times 10^{-22}$ eV (ULADM-1), $4.16\times 10^{-22}$ eV (ULADM-2) and $1.54\times 10^{-21}$ eV (ULADM-3), and late-forming DM models \cite{Das2011,Agarwal2015a} with transition redshift $z_t=5\times 10^5$ (LFDM-1), $8\times 10^5$ (LFDM-2) and $15\times 10^{15}$ (LFDM-3). Note that all the three ULADM models investigated here are in disagreement with the recent  constraints obtained from
the Lyman-$\alpha$ forest high redshift power spectrum which result in a lower limit at the 2$\sigma$ level of $\sim 2 \times 10^{-21}$ eV \cite{irsic17}, for a conservative analysis. However, given the intrinsic complementarity of the two observables, intensity mapping power spectrum probes the HI in mass while the forest probes the HI in volume, it is important to explore
these models to confirm or disproof the limits obtained with the forest (that are obtained at $z>3$).
These models are characterised by a suppression of the amplitude of matter density fluctuations at small scales below a characteristic length that for ULADM models depends on the particle mass, while in the case of LFDM models depends on the phase transition redshift. The halos in the simulation boxes are identified using the FoF algorithm \cite{pFoF}. Spurious artificial halos that form due to the sampling of numerical Poisson noise below the cut-off scale of the initial power spectrum of these models have been removed using the approach described in \cite{Agarwal2015b}. To be conservative we further retain only halos with 300 particles thus corresponding to a minimum halo mass of the catalogs of $M_{\rm min}^{27.5}=5\cdot 10^8$ M$_{\odot}$ $h^{-1}$. The cosmological model parameters have been set to those of a reference $\Lambda$CDM simulation ($\Lambda$CDM-S) of the same box-length and with equal number of particles with $\Omega_m=0.3$, $\Omega_b=0.046$, $h=0.7$, $\sigma_8=0.8$ and $n_s=0.99$.

In Fig.~\ref{fig2} we plot the ratio of the halo mass function of non-standard DM models relative to the reference $\Lambda$CDM simulation at $z=1$ (top panel) and $2$ (bottom panel). As expected these models exhibit suppressed halo abundances at low masses compared to the $\Lambda$CDM case, with the mass scale cut-off depending on the specificities of the underlying DM model. The larger the ULADM particles mass or equivalently the higher the phase transition redshift of LFDM models the lower the mass scale cut-off and the smaller the deviation from $\Lambda$CDM in the simulated mass range. By comparing the trends in Fig.~\ref{fig1} and Fig.~\ref{fig2} we can clearly see that non-standard DE and DM models, exhibit different mass dependent deviations from $\Lambda$CDM. As we will see these will contribute to having different predictions for the HI clustering signal.

\section{HI power spectrum of non-standard DE and DM models} \label{HIPk}
Our goal is to study the imprints of DE and DM models on the HI power spectrum. Since the HI distribution is modelled upon the results of numerical simulations, we first evaluate the impact of numerical effects due to the finite volume and the mass resolution of the halo catalogs.

\subsection{Minimum Halo Mass \& Volume Effects}

\begin{figure}[t]
 	\centering 
 	\includegraphics[width=.8\textwidth]{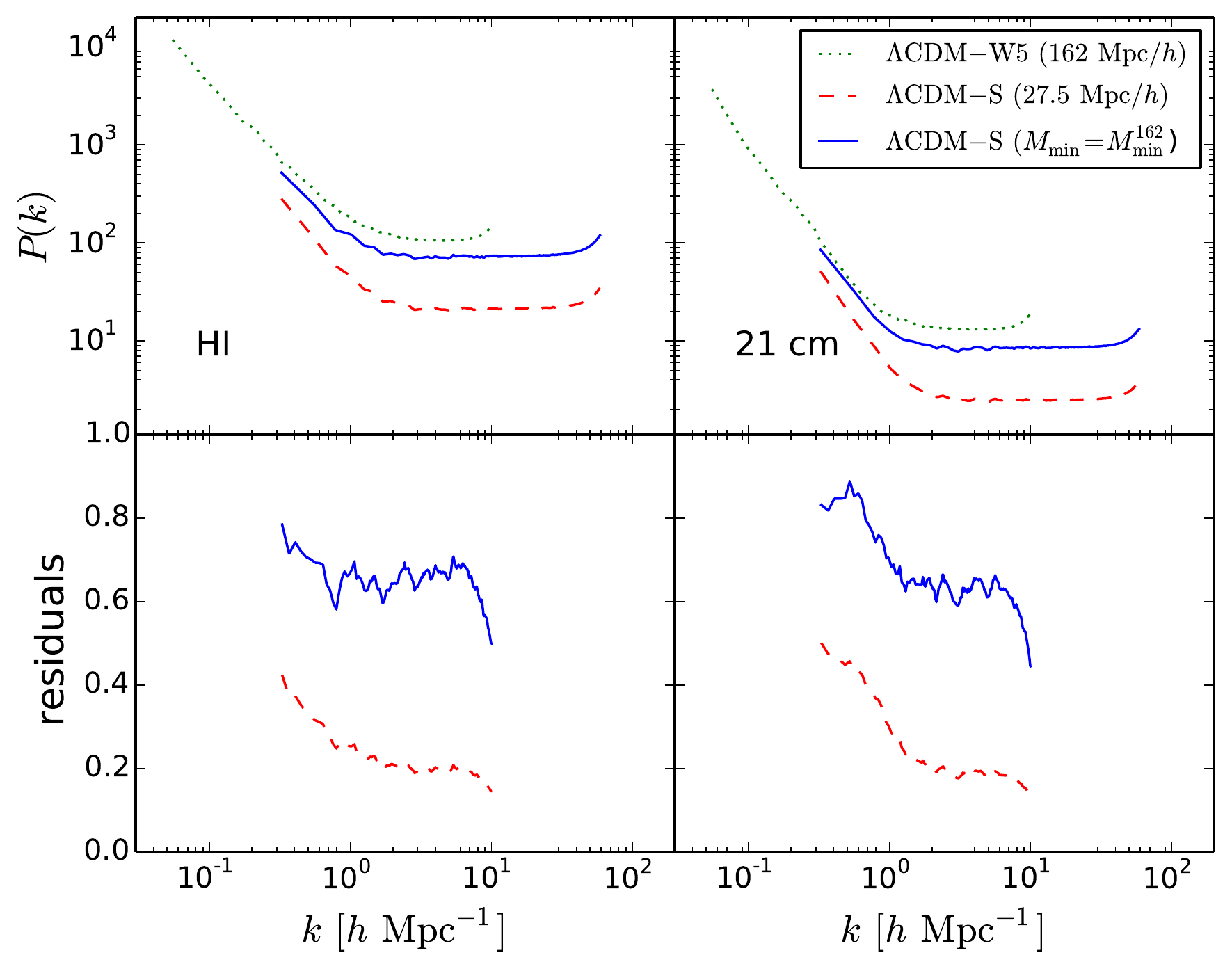}
 	\caption{\label{fig3} Power spectrum of HI (left panels) and of the 21cm signal (right panels) at $z=1$ for the different $\Lambda$CDM simulations: $\Lambda$CDM-W5 in green dotted line and the $\Lambda$CDM-S catalog having set $M_{\rm min}=M_{\rm min}^{27.5}$ (red dashed) and $M_{\rm min}=M_{\rm min}^{162}$ (blue solid). In the bottom panels the ratios of the $\Lambda$CDM-S power spectra over the $\Lambda$CDM-W5 one.}
 \end{figure}
 \begin{figure}[h]
 	\centering 
 	\includegraphics[width=.8\textwidth]{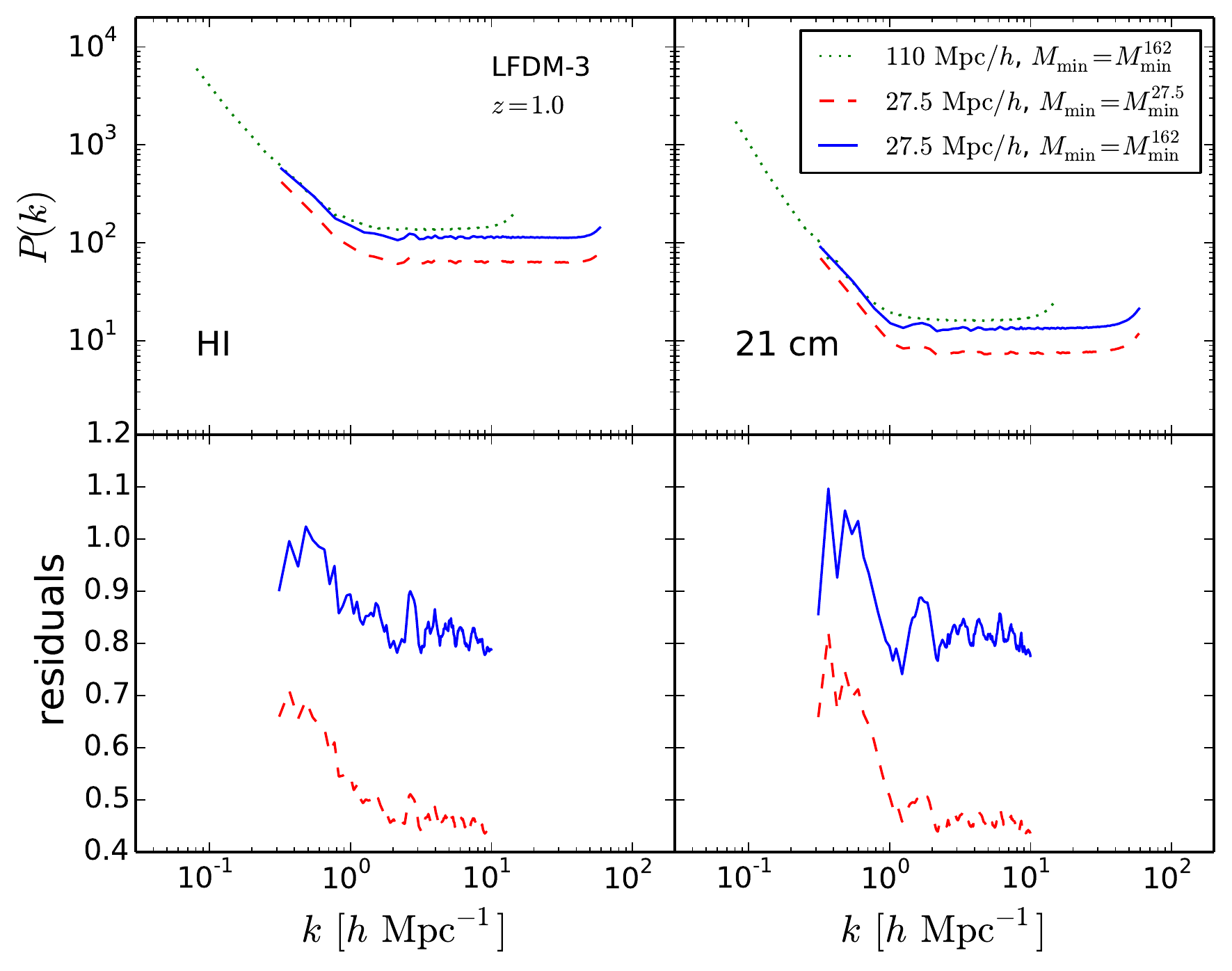}
 	\caption{\label{fig4} Power spectrum of HI (left panels) and of the 21cm signal (right panels) at $z=1$ of the LFDM-3 scenario for different simulation box size and value of $M_{\rm min}$ as in Eq.~(\ref{mhimodel}). In the bottom panels their ratios.}
 \end{figure}

As discussed in Sec.~\ref{HImodel}, the HI distribution model depends on the specification of a minimum halo mass containing the neutral gas cloud ($M_{\rm min}$). Although largely uncertain a conservative guess is to set $M_{\rm min}\sim 10^9$ M$_{\odot}$, however such a mass scale is resolved only in the case of the non-standard DM halo catalogs from the $27.5$ Mpc $h^{-1}$ box-length simulations with $1024^3$ particles. In the case of the dynamical DE model simulations, the minimum halo mass is about a factor 10 larger\footnote{Notice that this $M_{\rm min}$ value is anyway compatible with what found in \cite{castorina}, where it is tuned by requiring $M_{\rm HI}(M)$ to reproduce the observed bias of the Damped Lyman $\alpha$ systems at $z=2.3$.}. Hence, it is important to evaluate the impact of $M_{\rm min}$ on the HI power spectrum. 

In Fig.~\ref{fig3} we plot the HI power spectrum at $z=1$ for the $\Lambda$CDM-W5 model simulation (green dotted line) and from the $\Lambda$CDM-S catalog having set $M_{\rm min}=M_{\rm min}^{27.5}$ (red dashed line) and $M_{\rm min}=M_{\rm min}^{162}$ (blue solid line) in Eq.~(\ref{mhimodel}). The power spectrum from the larger simulation box covers a wider range of low-$k$ modes than the smaller box, the latter on the other hand extends to larger $k$ modes as the corresponding simulation has higher spatial and mass resolution. It is important to notice that in both cases the spectra exhibit an unphysical flattening of power at $k\gtrsim 2$ Mpc$^{-1}$ $h$, which is due to the lack of modelling the HI distribution within the halos (see appendix A in \cite{Carucci2017}). We can see that the HI power spectrum from the $27.5$ Mpc $h^{-1}$ box-length halo catalog increases in amplitude when increasing the value of $M_{\rm min}$. This is because the total amount of HI is fixed and by increasing $M_{\rm min}$ we assign more HI mass to more massive halos which are more clustered then low mass ones, thus leading to a larger amplitude of the HI power spectrum. Notice also that as we set $M_{\rm min}=M_{\rm min}^{162}$, the HI power spectrum of the $27.5$ Mpc $h^{-1}$ box-length halo catalog lies closer to that of the $\Lambda$CDM-W5 model with difference of $\sim 20-30\%$, however this cannot be attributed uniquely to a volume effect since the two $\Lambda$CDM simulations have slightly different values of the cosmological parameters.

Since we do not have a larger volume simulation of the $\Lambda$CDM-S model, we address this point using halo catalogs of the LFDM-3 model for which we have simulations of $27.5$ Mpc $h^{-1}$ box-length with $1024^3$ particles and an additional run with $110$ Mpc $h^{-1}$ box-length and $2048^3$ particles. The corresponding HI power spectra are shown in Fig.~\ref{fig4}. Again we notice a flattening of the power spectrum for $k \gtrsim 2\,h$ Mpc$^{-1}$. As in the previous case, we can see that increasing $M_{\rm min}$ from $M_{\rm min}^{27.5}$ to $M_{\rm min}^{162}$ leads to a larger amplitude of the HI spectrum by an amount similar to that found in $\Lambda$CDM-S case. Moreover, the HI spectrum from the $27.5$ Mpc $h^{-1}$ box-length catalog with $M_{\rm min}=M_{\rm min}^{162}$ (blue solid line) differs by less than $\lesssim 10\%$ from that of the $110$ Mpc $h^{-1}$ box-length simulation with $M_{\rm min}$ set to the same value (green dotted line). 

This analysis suggests that since numerical effects alter the HI power spectrum in the same way across different models, then by focusing on the relative difference of the HI spectra among these models is largely insensitive to volume and mass resolution effects.

As we do not model the HI gas distribution within halos, hereafter we only consider the HI spectra for $k \lesssim 2\,h$ Mpc$^{-1}$.
 
\subsection{Cosmological Dependence and Redshift Evolution of HI spectra}

\begin{figure}
	\centering
	\begin{subfigure}{.5\textwidth}
		\centering
		\includegraphics[width=0.9\linewidth]{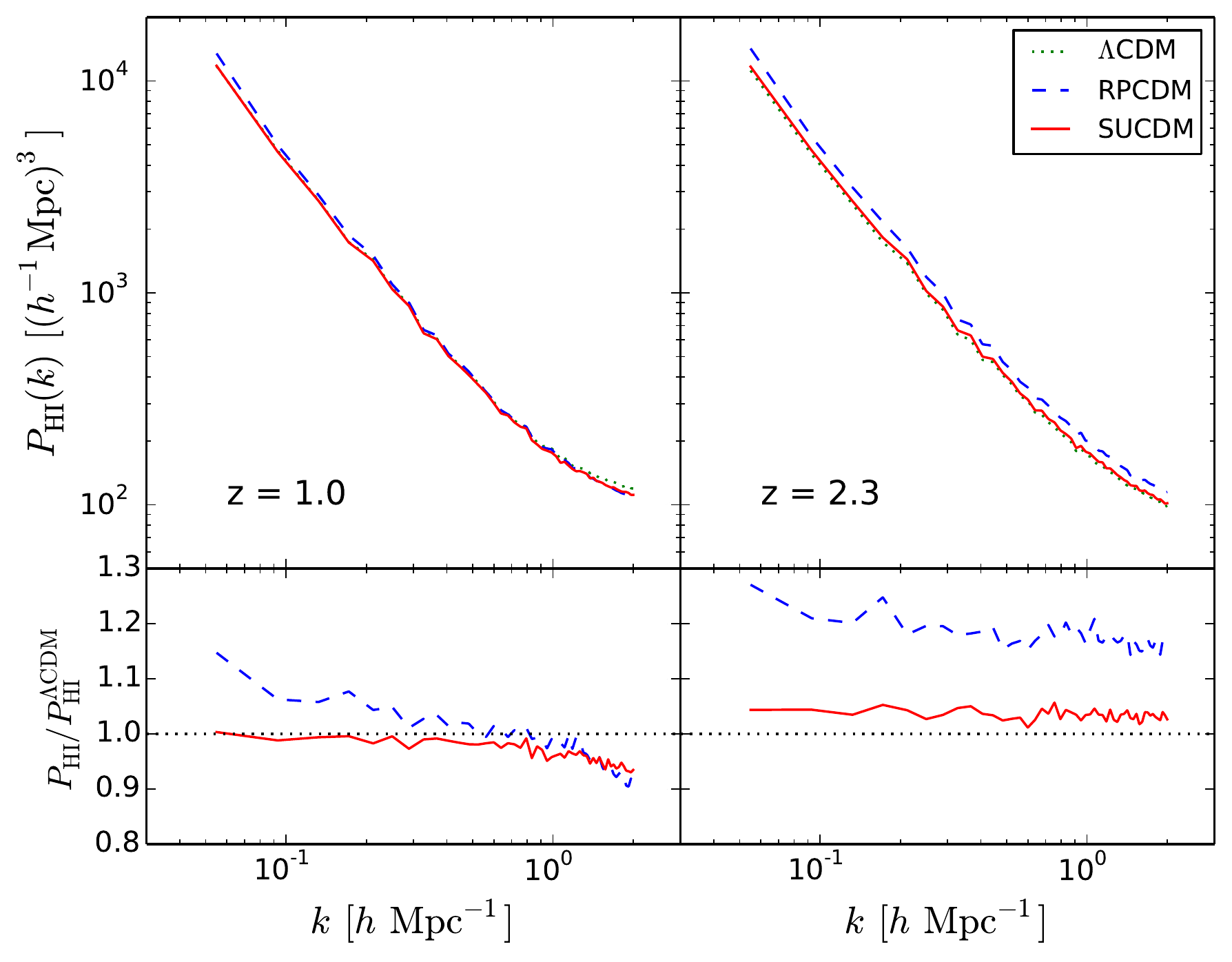}
		\caption{}
		\label{DE_HIspectra}
	\end{subfigure}%
	\begin{subfigure}{.5\textwidth}
		\centering
		\includegraphics[width=0.9\linewidth]{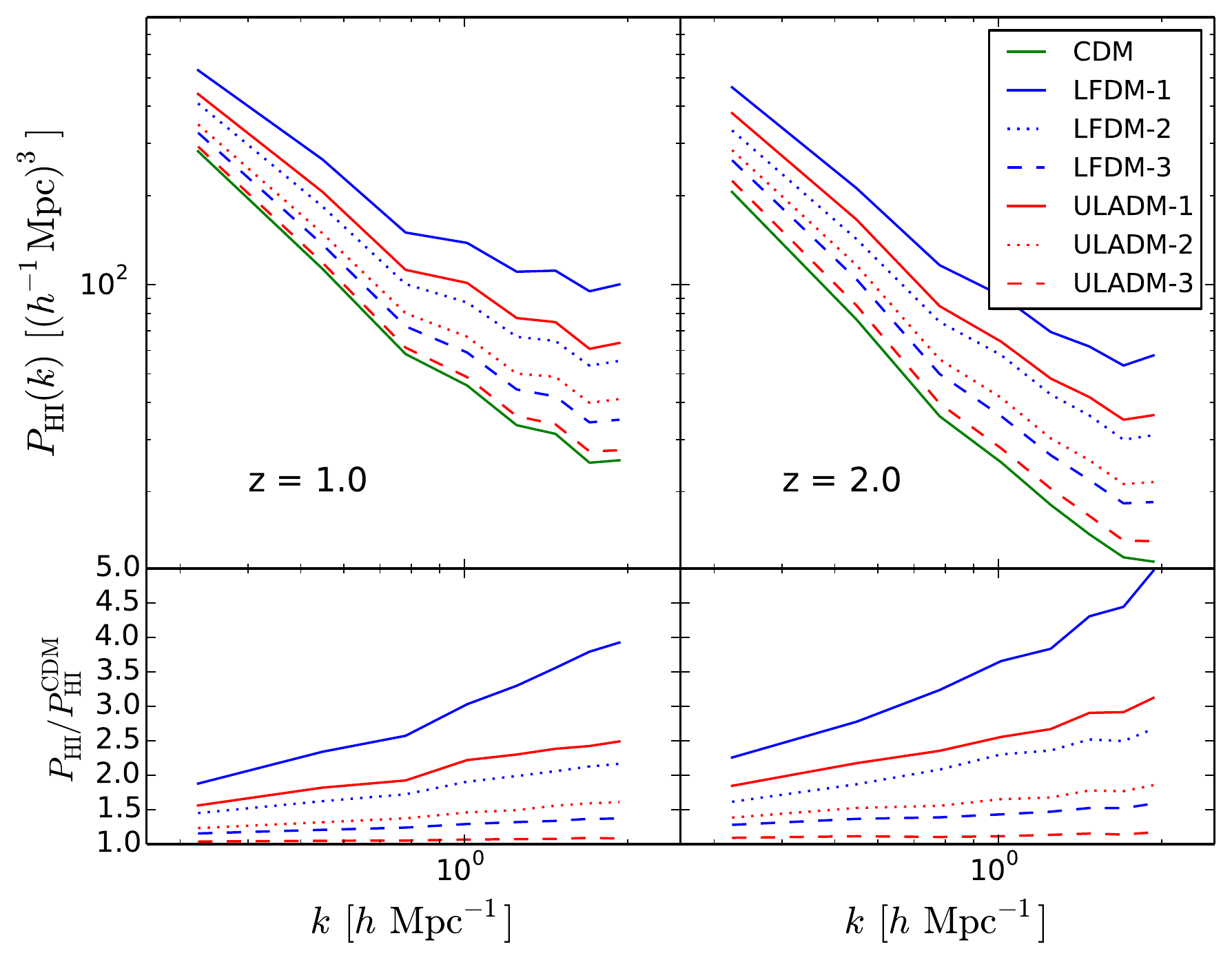}
		\caption{}
		\label{DM_HIspectra}
	\end{subfigure}
	\caption{HI power spectra (top panel) and relative difference with respect to the $\Lambda$CDM prediction (bottom panel) at $z=1$ and $2.3$ respectively, in the case of the non-standard DE (panel a) and DM (panel b) models respectively.}
	\label{figPkHI}
\end{figure}

We now focus on the signatures of the non-standard cosmological models on the HI spectra relative to the standard $\Lambda$CDM scenario.  

In Fig.~\ref{figPkHI} we plot the HI spectra for the DE (left panel) and DM (right panel) models at $z=1$ and $2.3$ respectively. In the bottom panels we show the relative differences with respect to the reference $\Lambda$CDM model. In the case of the non-standard DE models the differences among the HI spectra at large scales can be understood in terms of the evolution of the halo mass function. In fact, at $z=2$ the RPCDM and SUCDM models have spectra with amplitude larger than in the $\Lambda$CDM case. This is consistent with the fact that at this redshift the abundance of halos in the non-standard DE models are suppressed compared to the reference $\Lambda$CDM (see Fig.~\ref{fig1}), consequently since we distributed in the halo catalogs the same HI mass to all models, the HI  populates more massive halos in RPCDM and SUCDM than in the $\Lambda$CDM, as these are more biased the resulting clustering amplitude of the HI power spectrum is higher. At $z=1$, the halo abundance in the non-standard DE model is still suppressed compared to the $\Lambda$CDM, but higher than at $z=2.3$, hence the amplitude of the HI spectra decrease. However, notice that at small scales the SUCDM has less suppressed spectrum than the RPCDM. This is because of a competing effects due to the onset of highly non-linear clustering regime that depends on the linear growth of structures, which in the case of SUCDM is larger than RPCDM compared to the $\Lambda$CDM model (see \cite{Alimi2010}). 

For the non-standard DM models shown in the right panel of Fig.~\ref{figPkHI} the linear growth rate is identical to that of the reference $\Lambda$CDM model, thus the relative differences in the HI spectra at $z=1$ and $2.3$ are entirely due to the suppressed abundances for low halo masses (see Fig.~\ref{fig2}), which is also consistent with the findings of \citep{Carucci2015} on WDM models.

\section{21cm Intensity Map Power Spectra: SKA1-MID forecasts}\label{SKA}

\begin{figure}[t]
	\centering 
	\includegraphics[width=.85\textwidth]{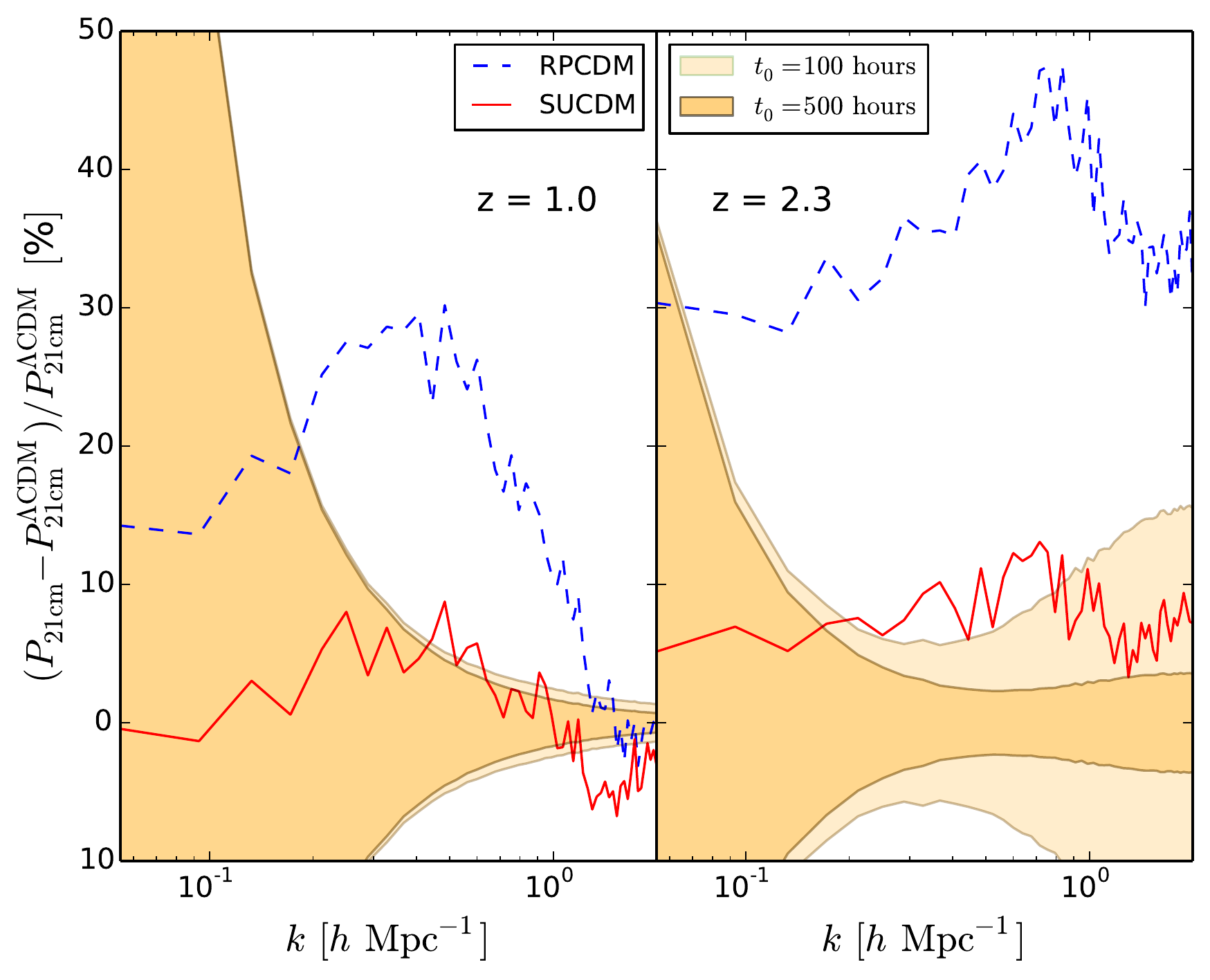}
	\caption{\label{SKA_de}  21cm intensity map power spectrum of the non-standard DE models relative to $\Lambda$CDM-W5 predictions at $z=1$ (left panel) and $2.3$ (right panel) respectively. The shaded area represents the expected errors from SKA1-MID measurements for the reference $\Lambda$CDM-W5 model,  $\sigma[P^{\Lambda{\rm CDM}}_{21{\rm cm}}(k)] / P^{\Lambda{\rm CDM}}_{21{\rm cm}}(k)$, assuming $t_0 = 100$ (light shaded area) and $t_0 = 500$ (dark shaded area) observing hours.}
\end{figure}

\begin{figure}
\centering 
\includegraphics[width=.85\textwidth]{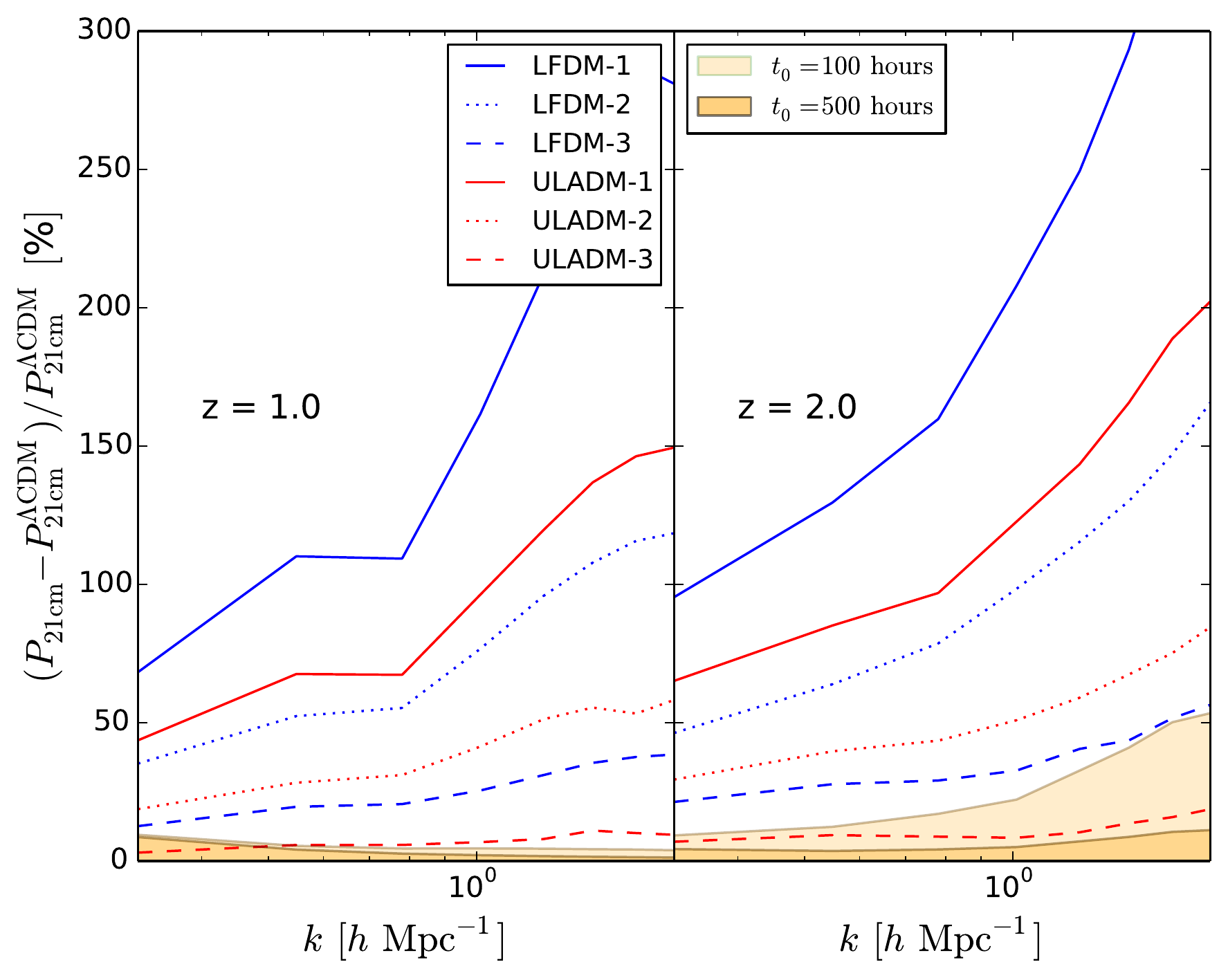}
\caption{\label{SKA_dm} As in Fig.~\ref{SKA_de} for the non-standard DM models relative to the reference $\Lambda$CDM-S model.}
\end{figure}

We now focus on the 21cm intensity map power spectrum. Differently from the HI spectra, the 21cm encodes additional cosmological information since it traces in redshift space the location of HI "clouds" whose peculiar velocities alter the clustering signal. In Fig.~\ref{SKA_de} we plot the relative difference of the 21cm power spectrum between the non-standard DE models and the reference $\Lambda$CDM-W5 at $z=1$ (left panel) and $z=2.3$ (right panel) respectively, while in Fig.~\ref{SKA_dm} we plot the spectra in the case of the non-standard DM models relative to the reference $\Lambda$CDM-S. 

It is worth noticing that the 21cm spectra of the DE models differ from the $\Lambda$CDM case not only by an amplitude factor but also on the scale dependence of the signal, in particular we may notice a change of the slope of the spectra at small scales. In the case of the non-standard DM models the differences with respect to the standard cosmological scenario show a very different trend, with differences increasing at small scales. This suggests that in principle the imprints of DE and DM models can be distinguished from one another through 21cm intensity map measurements. 

To be more quantitative, following the computation in \cite{Carucci2017} we have estimated the 1$\sigma$ errors on 21cm power spectrum measurements expected from the SKA1-MID radio telescope in interferometry for the reference $\Lambda$CDM models. These are shown in Fig.~\ref{SKA_de} and Fig.~\ref{SKA_dm} as shaded regions for $t_0 = 100$ (light shaded area) and $t_0 = 500$ (dark shaded area) observing hours respectively, having assumed  that astrophysical and atmospheric foreground contaminations and radio interferences have been removed from data. 

We can see that the RPCDM model can be distinguished from the $\Lambda$CDM at high-statistical significance. Even the SUCDM model, characterised by a cosmic expansion and a linear growth rate similar to that of the $\Lambda$CDM, can be potentially distinguished at more that $1\sigma$ at $z=2.3$ in the range of scales corresponding to $0.02\lesssim k\,[{\rm Mpc}^{-1}\,h]\lesssim 2$. Similarly, the non-standard DM models considered here should be detectable or ruled out with future SKA observations.

\section{Conclusion} \label{conclusions}
Future measurements of the clustering of matter in the Universe are a promising tool to infer new insights onto the properties of dark energy and dark matter. In particular, observations of the 21cm emission from HI  in distant galaxies can provide a complementary probe of matter clustering across a wide range of scales and at high redshifts. 

Here, we have studied the imprint of non-standard DE and DM models on the clustering of HI and 21cm power spectra. Using halo catalogs from N-body simulations we associate HI gas mass to halos using a simple model prescription to estimate the spatial distribution of HI  and infer the 21cm intensity map power spectra. 

We find that the simulated DE models leave a characteristic imprint on the HI spectra which differ from that of a reference $\Lambda$CDM model. In particular, although the models are statistically indistinguishable from one another using cosmological observations currently available, differences in the abundance of halos and the onset of the non-linear clustering regime impact significantly on the power distribution of the HI spectra between large and small scales at high-redshifts. The non-standard DM model considered here are also statistically consistent with large scale structure observations and only differ from the $\Lambda$CDM scenario for the suppressed abundance of low mass halos. This causes the HI power spectra to have larger amplitudes than the $\Lambda$CDM case and thus differ from the predictions of the dynamical DE models. Such model dependent features are also manifest in the 21cm intensity map spectra such that future SKA measurements should be able to detect or rule out the non-standard DE and DM models we have considered.

This work indicates that measurements of the 21cm signal can provide strong constraints on DE and DM models. However, as the result rely on a simply modelling of the HI cloud distribution on dark matter only simulations, it suggests to further pursue this investigation through the realisation of hydrodynamic simulations of the HI distribution in non-standard cosmological scenarios.

%\begin{figure}[tbp]
%\centering 
%\includegraphics[width=.85\textwidth]{compare_wwa_2D_params_H0.pdf}
%\caption{\label{fig:planckfig} .}
%\end{figure}

%\begin{table}[tbp]
%\centering
%\begin{tabular}{|cccc|}
%\hline
%$\omega_b$ & $\omega_m$ &  $n_s$ & $\sigma_8$ \\
%\hline
%0.0215/0.0235 & 0.120/0.155 & 0.85/1.05 & 0.6/0.9 \\
%\hline
%\end{tabular}
%\caption{\label{tab:tab1}.}
%\end{table}

%Non linear features are surely present also at smaller $k$'s, but a 
%linear expression is better approximate than $\sim 1$--$1.5\, \%$. 
%Spectral distorsions due to baryon effects are surely present also 
%for $k <\sim 2\, h{\rm Mpc}^{-1}$, although their significance 
%depends on cosmological parameters and the limits considered are 
%those for which {\it some models} exibit spectra distorted by $< 1$--
%$ 1.5\, \%$.
%\subsection{Auxiliary model definition}
%\appendix

\acknowledgments 
IPC thanks Francisco Villaescusa-Navarro for interesting discussions and suggestions. This work was granted access to the HPC resources of TGCC under the allocation 2016-042287 made by GENCI (Grand Equipement National de Calcul Intensif) on the machine Curie. The research leading to these results has received funding from the European Research Council under the European Community Seventh Framework Programme (FP7/2007-2013 Grant Agreement no. 279954). We acknowledge support from the DIM ACAV of the Region Ile-de-France. MV is supported by the ERC-StG cosmoIGM,
and PRIN MIUR, PRIN INAF and  INFN INDARK I.S. PD51 grants.


\begin{thebibliography}{99}
\bibitem{Riess1998} A. G. Riess et al., {\it Observational Evidence from Supernovae for an Accelerating Universe and a Cosmological Constant}, Astrophys. J. 116, 1009 (1998)
\bibitem{Perlmutter1999} S. Perlmutter et al., {\it Measurements of $\Omega$ and $\Lambda$ from 42 High-Redshift Supernovae}, Astrophys. J. 517, 565 (1999)
\bibitem{Tegmark2004} M. Tegmark et al., {\it Cosmological parameters from SDSS and WMAP}, Phys. Rev. D 69, 103501 (2004)
\bibitem{Clowe2006} D. Clowe et al., {\it A Direct Empirical Proof of the Existence of Dark Matter}, Astrophys. J. 648, L109 (2006)
\bibitem{Massey2007} R. Massey et al., {\it Dark matter maps reveal cosmic scaffolding}, Nature 445, 286 (2007)
\bibitem{Planck2015} Planck Collaboration: P.A.R. Ade et al., {\it Planck 2015 results. XIII. Cosmological parameters}, Astron. \& Astrophys. 594, 13 (2016)
\bibitem{Carroll2001} S. M. Carroll, {\it The Cosmological Constant}, Living Rev. Relativ. 4, 1 (2001)
\bibitem{Wetterich1988} C. Wetterich, {\it Cosmology and the fate of dilatation symmetry}, Nucl. Phys. B302, 668 (1988)
\bibitem{Ratra1988} B. Ratra, P.J.E. Peebles, {\it Cosmological consequences of a rolling homogeneous scalar field}, Phys. Rev. D 37, 3406 (1988)
\bibitem{Caldwell1998} R.R. Caldwell, R. Dave, P.J. Steinhardt, {\it Cosmological Imprint of an Energy Component with
General Equation of State}, Phys. Rev. Lett. 80, 1582 (1998)
\bibitem{Clifton2012} T. Clifton, P.G. Ferreira, A. Padilla, C. Skordis, {\it Modified Gravity and Cosmology}, Phys. Rep. 513, 1 (2012)
\bibitem{LSST} LSST Science Collaborations: P.A. Abell et al., {\it LSST Science Book}, arXiv:0912.0201
\bibitem{Euclid} R. Laureijs et al., {\it Euclid Definition Study Report}, arXiv:1110.3193
\bibitem{Wyithe2007} J.S.B. Wyithe, A. Loeb, P. M. Geil, {\it Baryonic Acoustic Oscillations in 21cm Emission: A Probe of Dark Energy out to High Redshifts}, Mont. Not. Roy. Astron. Soc. 383, 1195 (2008)
\bibitem{Chang2008}, T.-C. Chang, U.-L. Pen, J.B. Peterson, P. McDonald, {\it Baryon Acoustic Oscillation Intensity Mapping of Dark Energy}, Phys. Rev. Lett. 100, 091303 (2008)
\bibitem{Bull2015} P. Bull, P.G. Ferreira, P. Patel, M.G. Santos, {\it Late-time cosmology with 21cm intensity mapping experiments}, Astrophys. J. 803, 21 (2015)
\bibitem{Kohri2017} K. Kohri, Y. Oyama, T. Sekiguchi, T. Takahashi, {\it Elucidating dark energy with future 21cm observations at the epoch of reionization}, JCAP 02, 024 (2017)
\bibitem{Sitwell2014} M. Sitwell, A. Mesinger, Y.-Z. Ma, K. Sigurdson, {\it The Imprint of Warm Dark Matter on the Cosmological 21-cm Signal},  Mont. Not. Roy. Astron. Soc. 438, 2664 (2014)
\bibitem{Sekiguchi2014} T. Sekiguchi, H. Tashiro, {\it Constraining warm dark matter with 21cm line fluctuations due to minihalos}, JCAP 08, 007 (2014)
\bibitem{Carucci2015} I.P. Carucci, F. Villaescusa-Navarro, M. Viel, A. Lapi, {\it Warm dark matter signatures on the 21cm power spectrum: Intensity mapping forecasts for SKA}, JCAP 07, 047 (2015)
\bibitem{Weinberg2015} D.H. Weinberg et al., {\it Cold dark matter: Controversies on small scales}, PNAS, 112, 12249 (2015)
\bibitem{Casarini2009} L. Casarini, A.V. Macci\'{o}, S.A. Bonometto, {\it Dynamical dark energy simulations: high accuracy power spectra at high redshift}, JCAP 03, 014 (2009)
\bibitem{Alimi2010} J.-M. Alimi et al., {\it Imprints of Dark Energy on Cosmic Structure Formation. I. Realistic
Quintessence Models}, Mont. Not. Roy. Astron. Soc. 401, 775 (2010)
\bibitem{Jenning2010} E. Jennings, C.M. Baugh, R.E. Angulo, S. Pascoli, {\it Simulations of quintessential cold dark
matter: beyond the cosmological constant}, Mont. Not. Roy. Astron. Soc. 401, 2181 (2010)
\bibitem{Casarini2010} L. Casarini, {\it High-precision spectra for dynamical Dark Energy cosmologies from constant-w models}, JCAP 08, 005 (2010)

\bibitem{baldi2010} M. Baldi, V. Pettorino, G. Robbers and V. Springel, {\it Hydrodynamical N-body simulations of coupled dark energy cosmologies},  Mont. Not. Roy. Astron. Soc. 403, 1684 (2010)
\bibitem{baldi2011} M. Baldi, {\it Time-dependent couplings in the dark sector: from background evolution to non-linear structure formation},  Mont. Not. Roy. Astron. Soc. 411, 1077 (2011)

\bibitem{Bagla2010} J.S. Bagla, N. Khandai, K.K. Data, {\it HI as a probe of the large-scale structure in the post-reionization universe}, Mont. Not. Roy. Astron. Soc. 407, 567 (2010)
\bibitem{Carucci2017} I.P. Carucci, F. Villaescusa-Navarro, M. Viel, {\it The cross-correlation between 21cm intensity mapping maps and the Ly$\alpha$ forest in the post-reionization era}, JCAP 04, 001 (2017)
\bibitem{Villaescusa2015} F. Villaescusa-Navarro, P. Bull, M. Viel, {\it Weighing Neutrinos with with Cosmic Neutral Hydrogen}, Astrophys. J. 814, 146 (2015)
\bibitem{Villaescusa2016} F. Villaescusa-Navarro et al., {\it Neutral hydrogen in galaxy clusters: impact of AGN feedback and implications for intensity mapping}, Mont. Not. Roy. Astron. Soc. 456, 3553 (2016)
\bibitem{Pontzen2008} A. Pontzen et al., {\it Damped Lyman$\alpha$ systems in galaxy formation simulations}, Mont. Not. Roy. Astron. Soc. 390, 1349 (2008)
\bibitem{Noterdaeme2009} P. Noterdaeme, P. Petitjean, C. Ledoux, R. Srianand, {\it Evolution of the cosmological mass density of neutral gas from Sloan Digital Sky Survey II - Data Release 7}, Astron. \& Astrophys. 505, 1087 (2009)
\bibitem{Pochaska2009} J.X. Prochaska, A.M. Wolfe, {\it On the (Non)Evolution of HI Gas in Galaxies Over Cosmic Time}, Astrophys. J. 696, 1543 (2009)
\bibitem{SUGRA} P. Brax, J. Martin, {\it The Robustness of Quintessence}, Phys. Rev. D 61, 103502 (2000)
\bibitem{WMAP5} E. Komatsu et al., {\it Five-Year Wilkinson Microwave Anisotropy Probe Observations: Cosmological Interpretation}, Astrophys. J. Supp. 180, 330 (2009)
\bibitem{SN} M. Kowalkski et al., {\it Improved Cosmological Constraints from New, Old, and Combined Supernova Data Sets}, Astrophys. J. 686, 749 (2008)
\bibitem{Linder} E.V. Linder, {\it Exploring the Expansion History of the Universe}, Phys. Rev. Lett. 90, 091301
\bibitem{Chevalier} M. Chevallier, D. Polarski, {\it Accelerating Universes with Scaling Dark Matter}, Int. J. Mod. Phys. D 10, 213 (2001)
\bibitem{Courtin2011} J. Courtin et al., {\it Imprints of dark energy on cosmic structure formation: II) Non-Universality of the halo mass function},  Mont. Not. Roy. Astron. Soc. 410, 1911 (2011)
\bibitem{Marsh2016} D.J.E. Marsh, {\it Axion Cosmology}, Phys. Rep. 643, 1 (2016)
\bibitem{Das2011} S. Das, N. Weiner, {\it Late forming dark matter in theories of neutrino dark energy}, Phys. Rev. D 84, 123511 (2011)
\bibitem{Agarwal2015a} S. Agarwal, P.-S. Corasaniti, S. Das, Y. Rasera, {\it Small scale clustering of late forming dark matter}, Phys. Rev. D 92, 063502 (2015) 
\bibitem{Corasaniti2017} P.-S. Corasaniti, S. Agarwal. D.J.E. Marsh, S. Das, {\it Constraints on dark matter scenarios from measurements of the galaxy luminosity function at high redshifts}, Phys. Rev. D 95, 083512 (2017)
\bibitem{Agarwal2015b} S. Agarwal, P.-S. Corasaniti, {\it Structural properties of artificial halos in nonstandard dark matter simulations},  Phys.Rev. D 91, 123509 (2015) 
\bibitem{Mao2012} Y. Mao et al., {\it Redshift-space distortion of the 21-cm background from the epoch of reionization - I. Methodology re-examined},  Mont. Not. Roy. Astron. Soc. 422, 926 (2012)
\bibitem{pFoF} F. Roy, V. Bouillot, and Y. Rasera, {\it pFoF: a highly scalable halo-finder for large cosmological data sets},  Astron. Astrophys. 564, A13 (2014)


\bibitem{irsic17} V. Ir{\v s}i{\v c}, M. Viel, M.~G. Haehnelt, J.~S. Bolton, and G.~D. Becker, {\it First constraints on fuzzy dark matter from Lyman-$\alpha$ forest data and hydrodynamical simulations},  arXiv:1703.04683 (2017)

\bibitem{castorina} E. Castorina and F.  Villaescusa-Navarro, {\it On the spatial distribution of neutral hydrogen in the Universe: bias and shot-noise of the HI Power Spectrum},  arXiv:1609.05157 (2016)



\end{thebibliography}
\end{document}